\documentclass[aps,twocolumn,nofootinbib,showpacs,prd,aps,10pt]{revtex4}
\usepackage[dvips]{graphicx}
\usepackage[english]{babel}
\usepackage[T1]{fontenc}
\usepackage{mathrsfs}
\usepackage[tbtags]{amsmath}
\usepackage{amssymb}
\usepackage{amsxtra}
\usepackage{amsopn}
\usepackage{latexsym}
\usepackage[mathcal]{eucal}
\usepackage{mathtools}

\newcommand{\BE}{\begin{equation}}
\newcommand{\EE}{\end{equation}}
\newcommand{\BA}{\begin{align}}
\newcommand{\EA}{\end{align}}
\newcommand{\Tr}{\mathrm Tr}
\newcommand{\nn}{\nonumber}
\newcommand{\kkk}{ \frac{{\rm d}^4k}{(2\pi)^4}}
\newcommand{\qqq}{ \frac{{\rm d}^4q}{(2\pi)^4}}
\newcommand{\kkki}{ \frac{i{\rm d}^4k}{(2\pi)^4}}
\newcommand{\qqqi}{ \frac{i{\rm d}^4q}{(2\pi)^4}}
\newcommand{\kkkE}{ \frac{{\rm d}^4k_E}{(2\pi)^4}}
\newcommand{\qqqE}{ \frac{{\rm d}^4q_E}{(2\pi)^4}}

\begin{document}

\title{The gluon propagator in Feynman gauge by the method of stationary variance}

\author{Fabio Siringo}

\affiliation{Dipartimento di Fisica e Astronomia 
dell'Universit\`a di Catania,\\ 
INFN Sezione di Catania,
Via S.Sofia 64, I-95123 Catania, Italy}

\date{\today}
\begin{abstract}
The low-energy limit of pure Yang-Mills $SU(3)$ gauge theory is studied 
in Feynman gauge by the method of stationary variance, a genuine second-order 
variational method that is suited to deal with the minimal coupling of fermions in gauge theories. 
In terms of standard irreducible graphs, the stationary equations are written as a set of coupled 
non-linear  integral equations for the gluon and ghost propagators.
A physically sensible  solution is found for any strength of the coupling.
The gluon propagator is finite in the infrared, with a dynamical mass that decreases as a power
at high energies. At variance with some recent findings in Feynman gauge, the ghost
dressing function does not vanish in the infrared limit and a decoupling scenario emerges
as recently reported for the Landau gauge.
\end{abstract}



\maketitle

\section{introduction}

There is a growing consensus on the utility of variational methods as
analytical tools for a deeper understanding of the infrared (IR) limit
of non-Abelian gauge theories. The IR slavery of these theories makes
the standard perturbation theory useless below some energy scale, and
our theoretical knowledge of the IR limit relies on lattice simulation
and on non-perturbative techniques like functional renormalization group\cite{RG} 
and Dyson-Schwinger equations\cite{DSE}. Variational methods have been developed as
a complement to these analytical approaches, and their utility has been
proven by several authors in the last years\cite{kogan,reinhardt04,reinhardt05,reinhardt08,reinhardt11,
reinhardt14,szcz}.
Quite recently, the method of stationary variance\cite{sigma,sigma2} has been advocated as a powerful
second order extension of the Gaaussian Effective Potential (GEP)\cite{schiff,rosen,barnes,stevenson}. 
The GEP is a genuine variational method and has been successfully applied to many physical 
problems in field theory,
from scalar and electroweak theories\cite{stevenson,var,light,bubble,ibanez,su2,LR,HT} 
to superconductivity\cite{superc1,superc2,kim} and antiferromagnetism\cite{AF},
but turns out to be useless for gauge interacting fermions\cite{stancu2}.
Actually, since the GEP only contains first order terms, it is not suited
for describing the minimal coupling of gauge theories that has no first-order effects.
Several methods have been explored for including fermions\cite{HT} and higher order corrections\cite{stancu},
sometimes spoiling the genuine variational character of the method.

By a formal higher order extension of the GEP\cite{gep2} the method of stationary variance has been
developed as a genuine variational method that keeps in due account second order effects and seems
to be suited to deal with the minimal coupling of gauge theories. While the method has been shown to
be viable for the simple Abelian case of QED\cite{varqed}, its full potentialities have not been
explored yet. As a non-perturbative tool that can deal with fermions in gauge theories, the method
seems to be very useful for exploring the IR limit of QCD, and its natural application field is 
the non-Abelian $SU(3)$ gauge theory.

While a full study of QCD by that method is still far away, as a first step, in this paper we explore 
the solution of the stationary equations for  pure Yang-Mills $SU(3)$ theory. 
The method of stationary variance provides
a set of non-linear coupled integral equations whose solutions are the propagators for gluons and ghosts.
Therefore the work has a double motivation: the technical aim of showing that the method is viable and
a solution does exist (which is not obvious nor proven in general), and the physical interest on the gluon
propagator in the IR limit, where its properties seem to be related to the important issue of
confinement.

On the technical side, having shown that a sensible untrivial solution does exist is a major achievement that
opens the way to a broader study of QCD by the same method. Inclusion of quarks would be straightforward as some
fermions, the ghosts, are already present in the simple Yang-Mills theory, and they seem to play well their 
role of canceling the unphysical degrees of freedom. Other important technical issues are gauge invariance, 
renormalization and the choice of a physical scale.

The method is not gauge invariant, and we did not make any effort to restore gauge invariance at this stage.
There are several ways to attempt it\cite{kogan,aguilar10}, but in this first step we preferred to fix a gauge, 
namely the Feynman gauge where the calculation is easier, and explore the properties of the solution to see
if any unphysical feature emerges for the propagator and the polarization function. Actually the polarization
function is found approximately transverse up to a constant mass shift due to the dynamical mass generation.
As far as the solution satisfies, even approximately, the constraints imposed by gauge invariance, the method
is acceptable on the physical ground. On the other hand the gluon propagator is not a physical observable
and is known to be a gauge-dependent quantity. Of course, since the solution depends on the gauge, the
choice of working in Feynman gauge could be non-optimal, and the method could be improved by exploring
other gauge choices, like Landau gauge. Besides being easier, working in Feynman gauge is also 
interesting from the physical point of view, as there are very few data available on the gluon propagator in 
this gauge. 

Since lattice simulations are the most natural benchmark for any variational calculation in the IR limit,
we borrowed from lattice simulation the regulating scheme in terms of an energy cutoff and a 
bare coupling that depends on it. Renormalization Group (RG) invariance requires that the physical
observables are left invariant by a change of the cutoff that is followed by the corresponding change of 
the bare coupling. Then, renormalized physical quantities can be defined that do not depend on the cutoff.
The only free parameter of the theory is the energy scale, that must be fixed by a comparison with the
experimental data or lattice simulations. No other fit parameter has been introduced in the method.
Moreover, we do not need to insert any counterterm and especially mass counterterms that are forbidden 
by the gauge invariance of the Lagrangian.

On the physical side, the properties of the gluon propagator in Feynman gauge are basically unexplored.
In Coulomb gauge\cite{reinhardt04,szcz,reinhardt05,reinhardt08,reinhardt11}
and in Landau gauge\cite{reinhardt14,aguilar10,aguilar8,aguilar9,aguilar13,aguilar14,
aguilar14b,cucchieri2005,cucchieri2007,bogolubsky}
there has been an intense theoretical work in the last years.  In Landau gauge theoretical and
lattice data are generally explained in terms of a decoupling regime, with a finite
ghost dressing function and a finite massive gluon propagator. The more recent findings confirm the prediction
of a dynamical mass generation for the gluon\cite{cornwall82}.
In Feynman gauge we do not expect a very different scenario. A finite ghost propagator has been 
recently proposed\cite{aguilarFeynman}, but there are no lattice data available that could confirm it.
That makes the study of the Feynman gauge more interesting.
In the present work no important differences are found with respect to the Landau gauge. 
A decoupling scenario emerges, with very
flat ghost dressing functions, flatter than expected, and a finite gluon propagator in the IR limit.
A dynamical mass is found that saturates at about $0.8$ GeV and decreases as a power in the high energy
limit. Unfortunately the quantitative predictions are biased by an approximate estimate of the energy scale
due to the lack of lattice data in Feynman gauge.

This paper is organized as follows:
in Section II the method is described in detail for the special case of  pure $SU(3)$ Yang-Mills theory; 
then in Section III the stationary equations are derived and written
in terms of standard irreducible Feynman graphs; a comparison with other recent variational approaches
is reported in Section IV where some aspects of the method are clarified;
in Section V  the renormalization scheme is discussed and the numerical solutions
are studied in great detail, comparing them with the available lattice data;
in Section V a second-order approximation is introduced, and the numerical solution 
is  proposed as a better approximation for the propagator;
finally, in Section VI the results are discussed and several lines for future work are
outlined. Details on the numerical calculation and explicit integral expressions for
the Feynman graphs are reported in the appendix.

\section{Setup of the Method}

The method of stationary variance\cite{sigma,sigma2} is a second order 
variational technique that is suited to describe gauge theories with a minimal 
coupling like gauge theories\cite{gep2,varqed}, where first
order approximations like the GEP do not add anything to the standard treatment of
perturbation theory\cite{stancu2}. The method has been discussed in some detail 
in Ref.\cite{gep2} and applied to QED in Ref.\cite{varqed}. Here we give the
main details for a non-Abelian theory like $SU(3)$.

Let us consider a pure Yang-Mills  $SU(3)$ gauge theory without
external fermions. The Lagrangian can be written as
\BE
{\cal L}={\cal L}_{YM}+{\cal L}_{fix}
\EE
where ${\cal L}_{YM}$ is the Yang-Mills term
\BE
{\cal L}_{YM}=-\frac{1}{2} \Tr\left(  \hat F_{\mu\nu}\hat F^{\mu\nu}\right)
\EE
and ${\cal L}_{fix}$ is a guage fixing term.
In terms of the gauge fields, the tensor operator $\hat F_{\mu\nu}$ reads
\BE
\hat F_{\mu\nu}=\partial_\mu \hat A_\nu-\partial_\nu \hat A_\mu
-i g \left[\hat A_\mu, \hat A_\nu\right]
\EE
where
\BE
\hat A_\mu=\sum_{a} \hat X^a A^a_\mu
\EE
and the generators of $SU(3)$ satisfy the algebra
\BE
\left[ \hat X^a, \hat X^b\right]= i f_{abc} \hat X^c
\EE
with the structure constants normalized according to
\BE
f_{abc} f_{dbc}= N\delta_{ad}
\EE
and $N=3$.
Quite generally, the gauge-fixing term can be taken as
\BE
{\cal L}_{fix}=-\frac{1}{\xi} \Tr\left[(\partial_\mu \hat A^\mu)(\partial_\nu \hat A^\nu)\right]
\EE
and the quantum effective action $\Gamma[A^\prime]$, as a function of the external background field
$A^\prime$ can be written 
\BE
e^{i\Gamma[A^\prime]}=\int_{1PI} {\cal D}_{A} e^{iS[A^\prime+A]} J_{FP}[A^\prime+A]
\label{path}
\EE
where $S[A]$ is the action, $J_{FP}[A]$ is the Faddev-Popov determinant
and the path integral represents a sum over one particle irreducible (1PI) graphs\cite{weinbergII}.
Since the gauge symmetry is not broken and we are mainly interested in the propagators,
in the present paper we will limit to the physical vacuum at $A^\prime=0$,
while a more general formalism can be developed for a full study of the vertex functions
by keeping $A^\prime\not=0$ in order to take the derivatives of the effective action\cite{bubble}.

The determinant $J_{FP}$ can be expressed as a path integral over ghost fields
\BE
J_{FP}[A]=\int {\cal D}_{\omega,\omega^\star} e^{iS_{gh}[A,\omega,\omega^\star]} 
\label{pathghost}
\EE
and the effective action can be written as
\BE
e^{i\Gamma}=\int_{1PI} {\cal D}_{A, \omega, \omega^\star} e^{iS_0[A, \omega, \omega^\star]}
e^{iS_I[A, \omega, \omega^\star]}
\label{pathI}
\EE
where of course, the total action is  
\BE
S_{tot}=S_0+S_I=\int  {\cal L}_{YM}{\rm d}^4x+\int {\cal L}_{fix}{\rm d}^4x + S_{gh}
\EE
but we have the freedom to split it in the two parts, the free action $S_0$ and the
interaction $S_I$, by insertion of trial functions\cite{gep2}.
We {\it define} the free action $S_0$ as
\begin{align}
S_0&=\frac{1}{2}\int A^{a\mu}(x) {D^{-1}}^{ab}_{\mu\nu}(x,y) A^{b\nu}(y) {\rm d}^4x{\rm d}^4y \nn \\
&+\int \omega^\star_a(x) {G^{-1}}_{ab}(x,y) \omega_b (y) {\rm d}^4x{\rm d}^4y
\label{S0}
\end{align}
where $D^{ab}_{\mu\nu}(x,y)$ and $G_{ab}(x,y)$ are unknown trial matrix functions.
The interaction then follows by difference 
\BE
S_I=S_{tot}-S_0 
\EE
and can be formally written as the sum of a two-point term and three local terms: the ghost vertex,
the three-gluon vertex and the four-gluon vertex respectively
\BE
S_I=S_2+\int{\rm d}^4x \left[ {\cal L}_{gh} + {\cal L}_3 +   {\cal L}_4\right].
\label{SI}
\EE


In detail, the two-point interaction term can be written as
\begin{widetext}
\BE
S_2=\frac{1}{2}\int A^{a\mu}(x) 
\left[{{D_0}^{-1}}^{ab}_{\mu\nu}(x,y)- {D^{-1}}^{ab}_{\mu\nu}(x,y)\right]
A^{b\nu}(y) {\rm d}^4x{\rm d}^4y 
+\int \omega^\star_a(x)
\left[{{G_0}^{-1}}_{ab}(x,y)- {G^{-1}}_{ab}(x,y)\right]
\omega_b (y) {\rm d}^4x{\rm d}^4y
\label{S2}
\EE
\end{widetext}
where $D_0$ and $G_0$ are the standard free-particle propagators for
gluons and ghosts and their Fourier transforms read
\begin{align}
{D_0}^{ab}_{\mu\nu} (p)&=-\frac{\delta_{ab}}{ p^2}\left[\eta_{\mu\nu}
+(\xi-1)\frac{p_\mu p_\nu}{p^2}\right]\nn\\
{G_0}_{ab} (p)&=\frac{\delta_{ab}}{ p^2}
\label{D0}
\end{align}
where $\eta_{\mu\nu}$ is the metric tensor. 
The three local interaction terms are
\begin{align}
{\cal L}_3&=-g  f_{abc} (\partial_\mu A^a_\nu) A^{b\mu} A^{c\nu}\nn\\
{\cal L}_4&=-\frac{1}{4}g^2 f_{abc} f_{ade} A^b_\mu A^c_\nu A^{d\mu}A^{e\nu}\nn\\
{\cal L}_{gh}&=-g f_{abc} (\partial_\mu \omega^\star_a)\omega_b A^{c\mu}.
\label{Lint}
\end{align}
The trial functions $G_{ab}$, $D^{ab}_{\mu\nu}$ cancel in the total action $S_{tot}$ which
is exact and cannot depend on them. Thus this formal decomposition holds for any arbitrary
choice of the trial functions, provided that the integrals converge.
Standard Feynman graphs can be drawn for this theory with the trial propagators $D^{ab}_{\mu\nu}$ and 
$G_{ab}$ that play the role of free propagators, and the vertices that can be read from the
interaction action $S_I$ in Eq.(\ref{SI}). As shown in Fig.1, we have two-particle vertices for gluons
and ghosts that arise from the action term $S_2$ in Eq.(\ref{S2}), while the local terms in Eq.(\ref{Lint})
give rise to three- and four-particle vertices.

While the effective action $\Gamma$ can be evaluated by perturbation theory order by order, as a sum of
Feynman diagrams, a genuine variational method can be established by  the functional derivative of
the effective potential with respect to the trial propagators, in order to fulfill some given stationary
conditions. However, as recently discussed\cite{gep2}, the stationary conditions can be written in terms
of self-energy graphs directly, without having to write the effective potential, by use of the standard methods of 
perturbation theory.
  
\begin{figure}[b] \label{fig:vertex}
\centering
\includegraphics[width=0.4\textwidth,angle=0]{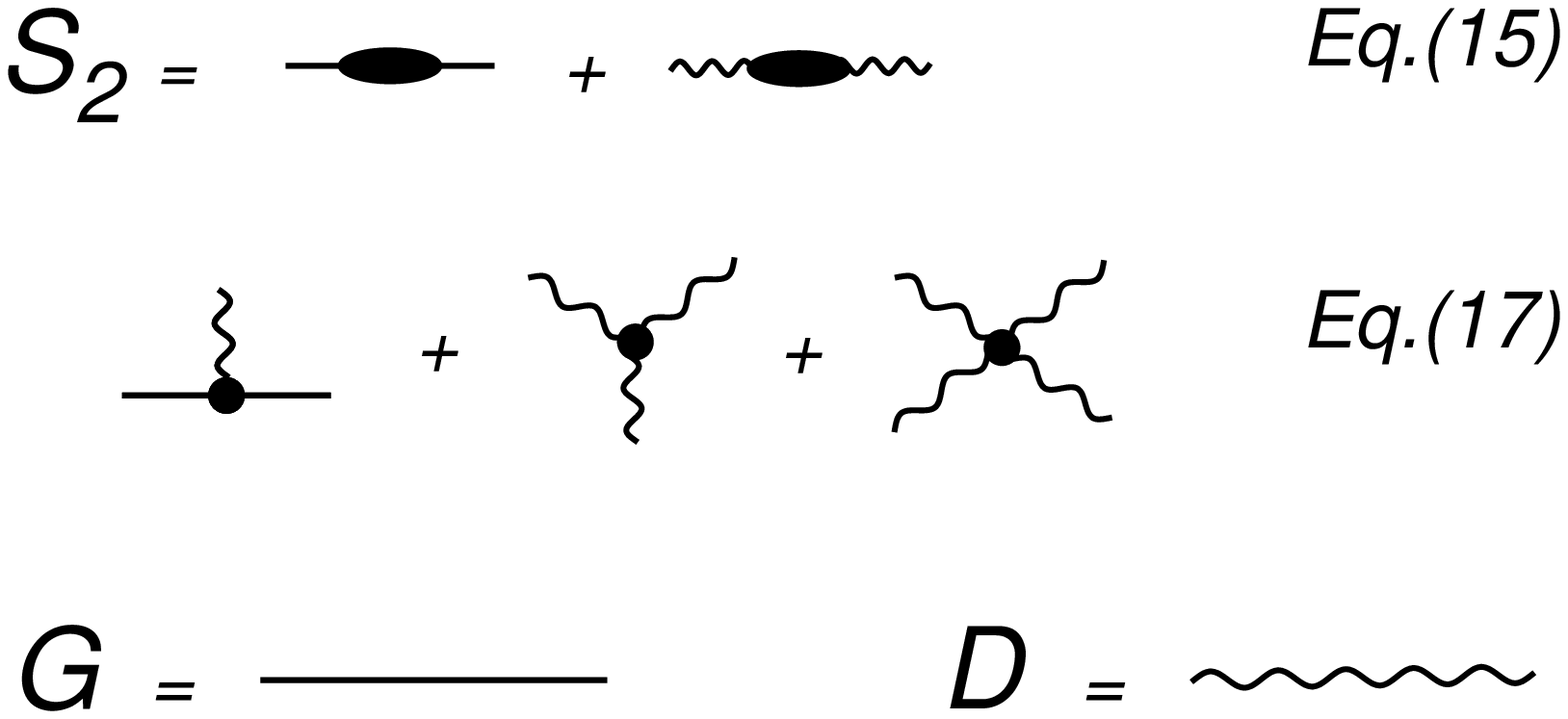}
\caption{The two-point vertices in the interaction $S_2$ of Eq.(\ref{S2}) are shown 
in the first line. The ghost vertex and the three- and four-gluon vertices of Eqs.(\ref{Lint})
are shown in the second line. In the last line the ghost (straight line) and gluon (wavy line)
trial propagators are displayed.}
\end{figure}

In this paper we test the method of stationary variance\cite{sigma,sigma2} that has been
shown to be viable in simple Abelian gauge theories like QED\cite{varqed}. 
According, the self-energy graphs are required up to second order, since the equation for stationary variance 
can be derived by the general connection that has been proven in Ref.\cite{gep2} 
\BE
\frac{\delta V_n}{\delta D_{\mu\nu}^{ab} (p)}=\frac{i}{2} 
\left( \Pi^{\nu\mu, ba}_n (p)-\Pi^{\nu\mu, ba}_{n-1}(p)\right),
\label{delVnP}
\EE
\BE
\frac{\delta V_n}{\delta G_{ab} (p)}=-i \left( \Sigma^{ba}_n (p)-\Sigma^{ba}_{n-1}(p)\right),
\label{delVnS}
\EE
where the nth-order gluon polarization function $\Pi_n^{\mu\nu, ab}$ and the nth-order ghost self-energy 
$\Sigma_n^{ab}$  are the sum of all nth-order connected two-point graphs  without tadpoles, 
while $V_n$ is the nth-order term of the effective
potential. First and second order two-point graphs are shown in  Fig.2.

\begin{figure}[t] \label{fig:sigma}
\centering
\includegraphics[width=0.29\textwidth,angle=-90]{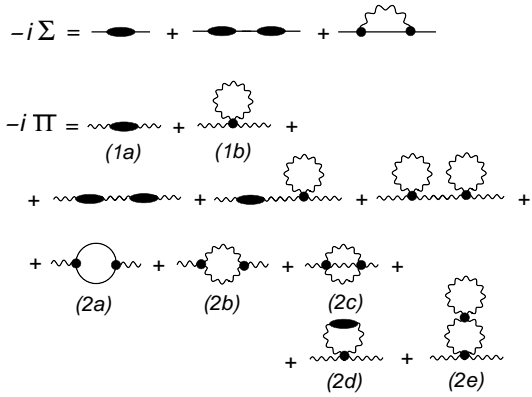}
\caption{First and second order two-point graphs contributing to the ghost self energy and the gluon
polarization. Second order terms include non-irreducible graphs.}
\end{figure}

For $n=2$ the second order term $V_2$ is the variance and its stationary conditions follow as
\begin{align}
 \Pi^{\nu\mu, ab}_2 (p)&=\Pi^{\nu\mu, ab}_1(p)\nn\\
 \Sigma^{ba}_2 (p)&=\Sigma^{ba}_1(p).
\label{stat}
\end{align}
These are the general stationary conditions that we will use in this paper.

The choice of Feynman gauge, $\xi=1$, simplifies the calculation once we
take
\BE
D^{ab}_{\mu\nu} (p)=\delta_{ab}\eta_{\mu\nu} D(p)=\delta_{ab}\eta_{\mu\nu}\frac{f(p)}{-p^2}
\label{D}
\EE
where $D(p)$  is an unknown trial function and $f(p)$ is a trial gluon dressing function.
That choice is perfectly legitimate, but is equivalent to a variation of the
trial propagator inside a more limited class of functions.
More generally,
if we take
\BE
D^{ab}_{\mu\nu} (p)=\delta_{ab} t_{\mu\nu} (p) D(p)
\label{Dgen}
\EE
where $t_{\mu\nu}$ is a given gauge dependent tensor,
the functional derivative can be written as
\BE
\frac{\delta}{\delta D(p)}= \sum_{ab,\mu\nu}\int\kkk \frac{\delta D^{ab}_{\mu\nu} (k)}{\delta D(p)}
\frac{\delta}{\delta D^{ab}_{\mu\nu}(k)}
\EE
and by Eq.(\ref{Dgen}) we can write
\BE
\frac{\delta}{\delta D(p)}= \sum_{ab,\mu\nu} \delta_{ab} t_{\mu\nu} (p)
\frac{\delta}{\delta D^{ab}_{\mu\nu}(p)}.
\EE
Thus, if we define the summed quantity
\BE
\Pi_n (p)= \frac{1}{4(N^2-1)} \sum_{ab, \mu\nu} \delta_{ab} t_{\mu\nu} (p) \Pi_n^{\mu\nu, ab}(p)
\label{Psum}
\EE
and insert it in Eq.(\ref{delVnP}), the functional derivative of $V_2$ with respect to $D(p)$ yields
the simple stationary equation
\BE
\Pi_2 (p)=\Pi_1(p)
\EE
which replaces the first of Eqs.(\ref{stat}).

In this paper we will limit to the special case of Feynman gauge and
take $t_{\mu\nu}=\eta_{\mu\nu}$ in the calculation. 
An interesting alternative would be
the choice of the Landau gauge, $\xi=0$.
In any case, color symmetry ensures that we can always take
\BE
G_{ab} (p)=\delta_{ab} G(p)=\delta_{ab}\frac{\chi (p)}{p^2}
\EE
where $\chi(p)$ is a trial ghost dressing function.

Despite their simple shape, the stationary equations contain all the one- and
two-loop graphs displayed in Fig.2, and are a set of coupled non-linear
integral equations for the trial functions $D$, $G$. It is not obvious
that a solution does exist, but we will show in the next sections that 
a solution can be found by a numerical integration.

\section{Stationary equations}

Before attempting a numerical solution of the stationary equations it is useful to
write them in more detail, in terms of proper (1PI) and reducible graphs.

The first order terms can be easily written as a sum of the first-order graphs of
Fig.2. We have a single tree graph $\Sigma_1$ for the ghost self energy (the first graph in Fig.2)
and making use of the explicit form of the vertices in the interaction  Eq.(\ref{SI}) 
we can write
\BE
-i \Sigma^{ab}_1(p)=i\delta_{ab}\left[ p^2- G^{-1}(p) \right]
\label{S1}
\EE

The first-order gluon polarization has a tree graph $\Pi_{1a}$ and a one-loop term 
$\Pi_{1b}$ as shown in the second line of Fig.2 
\begin{align}
-i \Pi^{\mu\nu, ab}_{1a}&=i\delta_{ab}\eta^{\mu\nu}\left[{D_0}^{-1}-D^{-1}\right] \nn\\
-i \Pi^{\mu\nu, ab}_{1b}&=i\delta_{ab}\eta^{\mu\nu}(3Ng^2) I_0^{(1)}
\label{P1}
\end{align}
where, according to Eq.(\ref{D0}), ${D_0}^{-1}(p)=-p^2$ and the integrals $I_n^{(m)}$ are
constant terms that in the Euclidean formalism can be written in terms of the dressing function as
\BE
I_n^{(m)}=\int\kkkE \frac{[f(k_E)]^m}{(k_E^2)^{n+1}}.
\label{Imn}
\EE
We assume that these integrals are made finite by a regulating scheme as discussed below.
Details on the calculation of this and all the other loop graphs of Fig.2 are given in the appendix.
The sum in Eq.(\ref{Psum}) is trivial and yields
\BE
\Pi_1=D^{-1}-\Delta^{-1}
\label{P1s}
\EE
where we have defined a renormalized zeroth-order massive propagator $\Delta$ as
\BE
\Delta (p)=\frac{1}{-p^2+M^2}
\EE
and the mass term $M^2$ is defined by the gap equation
\BE
M^2=3Ng^2 I_0^{(1)}=3Ng^2 \int\kkkE D(k_E).
\label{gap}
\EE

We observe that, as a first order approximation, the gap equations of the GEP are 
equivalent\cite{gep2} to the self-consistency conditions $\Pi_1=0$ and $\Sigma_1=0$ that
yield the simple decoupled analytical solution
\begin{align}
G(p)&=\frac{1}{p^2}\nn\\
D(p)&=\Delta (p)
\label{GEP}
\end{align}
with free propagators for ghosts and massive gluons. 
 
The second order terms require the sum of all the other graphs displayed in Fig.2. It is useful to introduce
a proper polarization function $\Pi_2^\star$ and a proper self energy $\Sigma_2^\star$, that are defined as 
the sum of second-order 1PI graphs, i.e. the last graph of the first line and the last five graphs at the bottom
respectively in Fig.2.
Assuming a sum over all indices according to Eq.(\ref{Psum}) and
recalling the diagonal matrix structure of first order terms in Eqs.(\ref{S1}),(\ref{P1}) 
we can write the second order functions as
\begin{align}
\Pi_2&=\Pi_2^\star+(\Pi_1)^2D\nn\\
\Sigma_2&=\Sigma_2^\star+(\Sigma_1)^2G
\end{align}
where the second term on the right hand side is the sum of the reducible graphs, and we are using the
obvious notation $\Sigma_n^{ab}=\delta_{ab}\Sigma_n$.
Inserting Eq.(\ref{P1s}) and Eq.(\ref{S1}), the stationary  conditions of Eq.(\ref{stat})  now take
the following form
\begin{align}
G(p)&=\frac{1}{ p^2}-\frac{\Sigma_2^\star(p)}{ p^4}\nn\\
D(p)&=\Delta(p)-[\Delta(p)]^2\Pi_2^\star(p).
\end{align}
Thus we only need to consider the 1PI graphs contributing to the proper second-order functions.
This pair of coupled non-linear integral equations is well suited for an iterative numerical
solution. They can be written in terms of the dressing functions, and switching to the Euclidean
formalism we can write them as
\begin{align}
\chi(p_E)&=\left[1+\frac{1}{ p_E^2}\Sigma_2^\star(p_E)\right]\nn\\
f(p_E)&=\frac{p_E^2}{p_E^2+M^2}\left[1-\frac{\Pi^\star_2(p_E)}{p_E^2+M^2}\right].
\label{dress}
\end{align}
Of course an iterative solution of these equations requires a numerical evaluation of the one-
and two-loop graphs contributing to the second-order proper functions $\Pi_2^\star$, $\Sigma_2^\star$
that we need at each step as functionals of the unknown trial dressing functions $f$, $\chi$. The details
on the numerical evaluation of the graphs are reported in Appendix A.

\section{Comparison with other variational methods}

Before going to the detail of the numerical solution, we would like to compare the formal results of the previous
sections with other variational methods that have been proposed.

The GEP is probably the simplest variational approach and it gives a dynamical mass generation for
the gluon as shown in Eqs.(\ref{GEP}). Moreover the same result cannot be obtained by perturbation
theory and is a genuine non-perturbative result. In fact, the mass term $M^2$ comes out from the
one-loop tadpole graph $\Pi_{1b}$ that vanishes in dimensional regularization when evaluated by
inserting the zeroth-order gluon propagator: perturbation theory cannot predict a finite mass at any
order. On the other hand, the first order stationary conditions of the GEP require a self-consistent
solution with a mass that is evaluated by the gap equation, Eq.(\ref{gap}), and can be written as
\BE
M^2=3Ng^2 \int_\Lambda\kkkE \frac{1}{k_E^2+M^2}
\label{GEPgap}
\EE
where $\int_\Lambda$ means that the integral has been regularized by a suitable cutoff $\Lambda$,
like in non-perturbative lattice calculations. While the integral does not vanish even in
dimensional regularization, the simple cutoff regularization seems to be more suited for a direct
comparison with lattice calculations.

Among the shorthands of the GEP we mention the constant mass, which does not decrease at large moments,
and mainly the known difficulties for dealing with fermions\cite{stancu,HT,gep2} like quarks and even ghosts.
In fact, according to Eqs.(\ref{GEP}), the ghosts are decoupled and do not play any role in the GEP.

Recently, a technique has been developed for including untrivial effects of the fermions in the GEP\cite{AF,HT},
and has been used for a non-perturbative study of the Higgs-top sector of the standard model\cite{HT}.
The technique, that can be seen as an improvement of RPA, was tested in the two-dimensional half-filled Hubbard 
model, predicting the correct antiferromagnetic limit in the strong coupling limit\cite{AF}.
It is instructive to see how the technique would improve the GEP, allowing for a correct inclusion of
the ghosts. It is based on an exact formal integration of fermions, yielding a pure bosonic
effective action. The action is then expanded in powers of the bosonic field, and the expansion is
eventually truncated at some order, yielding an approximate action that can be dealt with by a variational
method like the GEP. It is quite obvious that truncation spoils the approximation that ceases to be a genuine
variational approximation (it is well known that RPA is not a variational approximation).
In the present context of a pure $SU(3)$ theory, the exact integration of ghosts gives back the Faddev-Popov
determinant 
\BE
J_{FP}(A)=e^{i S_{eff}(A)}
\EE
that {\it defines} the effective action
\BE
S_{eff}(A)=-i \log J_{FP}(A).
\EE
On the other hand $J_{FP}$ is the determinant of the matrix ${\cal F}^{ab}(x,y)$ 
that can be formally written\cite{weinbergII}
\BE
{\cal F}=G_0^{-1}\cdot\left[1+G_0\cdot B\right]
\EE
where the matrix $B^{ab}(x,y)$ is the ghost-gluon vertex in ${\cal L}_{gh}$ that can
be written as
\BE
B^{ab}(x,y)=g f_{abc} \partial^\mu A_\mu^c(x)\delta^4 (x-y).
\EE
Then following Ref.\cite{HT}, the effective action admits the exact expansion
\BE
S_{eff}(A)=S_{eff} (0)-i \Tr\sum_{n=1}^{\infty}\frac{(-1)^{n+1}}{n}\left[G_0\cdot B\right]^n.
\label{exp}
\EE
All these terms must be added to the interaction $S_I$ of Eq.(\ref{SI}) and must be regarded
as {\it first} order vertices to be inserted in the evaluation of the first order polarization $\Pi_1$.
Then the simple first-order self consistency condition $\Pi_1=0$ suffices for determining the gap
equation of the improved GEP.
Of course, some truncation of the expansion is required in order to have a viable calculation scheme,
and the truncation spoils the accuracy of the variational method.
In the present context the technique turns out to be equivalent to the expansion proposed by Reinhardt
and Feuchter\cite{reinhardt05} and recently used in the Lagrangian formalism in Ref.\cite{reinhardt14}.
In more detail, the first untrivial term of the expansion is the quadratic one, yielding the 
correction $\delta S_I$ 
\BE
\delta S_I=\frac{i}{2}\Tr\left[G_0\cdot B\cdot G_0\cdot B\right].
\EE
Since $B$ is linear in the field $A$, this interaction term is quadratic, and gives rise to a two-point vertex.
By inspection, this composite vertex contains a loop of two ghost propagators connected by two ghost-gluon vertices, 
and its corresponding first order tree term in the polarization is just the second order 1PI ghost loop
$\Pi_{2a}$ which is displayed in Fig.2. Thus the improved GEP stationary equation $\Pi_1=0$ now
reads
\BE
\Pi_1=D^{-1}-\Delta^{-1}+\Pi_{2a}=0
\EE
and gives a massive propagator~\footnote{More generally, at first order there is no need to take the special 
matrix form Eq.(\ref{Dgen}) and the polarization $\Pi_{2a}$ can be regarded as a matrix. In that case, provided
that $\Pi_{2a}$ is replaced by $\Pi^\prime_{2a}$ as defined in Eq.(\ref{Piprime}), this massive propagator gives
just the coefficient of $\eta_{\mu\nu}$ which is the physically relevant part of the propagator. As shown by
Eq.(\ref{curv}) in the appendix, $-\Pi^\prime_{2a}$ is formally equal to the curvature function of 
Ref.\cite{reinhardt14}.} 
\BE
D(p_E)=[p_E^2+ \Omega^2(p_E)]^{-1}
\label{curvature1}
\EE
with a mass $\Omega(p)$ that depends on $p$ and is given by the modified gap equations
\BE
\Omega^2 (p_E)= M^2-\Pi_{2a} (p_E)
\label{curvature2}
\EE
\BE
M^2=3Ng^2 \int_\Lambda\kkkE \frac{1}{k_E^2+\Omega^2(k_E)}
\label{GEPgap2}
\EE
where the ghost loop $\Pi_{2a}$ plays the role of the curvature function of Ref.\cite{reinhardt14}
as shown by Eq.(\ref{curv}) in the appendix.
In that work the approximation is improved by including an infinite class of higher order terms in the
expansion Eq.(\ref{exp}). That can be formally done by substituting a dressed ghost propagator $G$ for the
bare one $G_0$ in the ghost loop, that is equivalent to sum up an infinite series of higher order
graphs. As shown in that work the self consistency of the dressed propagator has important effects on
the gluon propagator. That seems to be a limit of the simple GEP, while the second order method of
stationary variance yields coupled self-consistency equations, Eqs.(\ref{dress}), for the ghost and gluon 
dressing functions.

A final note on the differences between the GEP and the present method of stationary variance comes
from a comparison of the higher-order propagators.
We can regard the trial propagators as the starting point of an optimized perturbation theory, and then
write higher-order Feynman graphs for the propagators. By Dyson equations, the nth-order propagator $D_{(n)}$ 
follows from the nth-order proper polarization $\Pi_{(n)}^\star$ as~\footnote{We denote by  $X_{(n)}$ the {\it total}
nth-order value of $X$ while $X_n$ is the single nth-order term: $X_{(n)}=\sum_{i=0}^n X_i$.
With the same notation $D\equiv D_{(0)}$ is the zeroth-order approximation.}
\BE
D_{(n)}^{-1}=D^{-1}-\Pi_{(n)}^\star
\label{Dn}
\EE
where in general, for $n>1$ the polarization $\Pi^\star_{(n)}$ is a matrix, but we omit the indices for brevity.
For the GEP, since $\Pi_1=0$ we obtain the self-consistency condition $D_{(1)}=D$.
This property is lost at higher orders: in the present scheme of the stationary variance the
total second-order proper polarization is
\BE
\Pi^\star_{(2)}=\Pi_1+\Pi_2^\star=D^{-1}-\Delta^{-1}+\Pi_2^\star
\EE
and the second order propagator reads
\BE
D_{(2)}^{-1}=\Delta^{-1}-\Pi_2^\star=p_E^2+M^2-\Pi_2^\star(p_E)
\label{D2}
\EE
A comparison with Eqs.(\ref{curvature1}),(\ref{curvature2}) shows that the second order
gluon propagator extends the improved GEP\cite{AF,HT} or the curvature approximation of Ref.\cite{reinhardt14}
by substituting the whole second order proper self-energy $\Pi^\star_2$ for the single ghost loop $\Pi_{2a}$
(the curvature of Ref.\cite{reinhardt14}). By itself that does not imply a better approximation, but 
we expect a richer description at least, and an improvement of gauge invariance as the graphs $\Pi_{2a}$ and
$\Pi_{2b}$ are now summed together as they should.

\section{Regularization and numerical solutions}

The method of stationary variance provides a set of coupled non-linear integral
equations for the dressing functions. However, there is no proof that the stationary conditions
Eqs.(\ref{dress}) have any solution at all. Actually, for any choice of the bare coupling $g$,
Eqs.(\ref{dress}) can be iterated and show a fast convergence towards a stable solution.
The existence of a stable and physically reasonable solution for the method of stationary variance
is one of the main achievements of the present paper, since the method can be developed further
as a non-perturbative tool for the study of QCD. An analytical proof of existence was only
given before under some special constraints and for the simpler case of an Abelian gauge theory\cite{varqed}.

For a numerical solution of the coupled set of stationary conditions, Eqs.(\ref{dress}),
we first need to regularize all the diverging integrals that are reported in detail in the appendix.
Dimensional regularization does not seem to be the best choice for a non-perturbative variational
approach because of the unknown form of the trial propagators that would require a spectral 
representation as in Ref.\cite{varqed}. Moreover, we cannot rely on a perturbative
renormalization, order by order, but rather we should consider a non-perturbative multiplicative
renormalization scheme. Since the variational method is not gauge invariant, the regulator can even 
break gauge symmetry, as we expect that gauge invariance should be recovered in physical observables
only approximately in the present approximation. In that respect the gauge parameter $\xi$ could
even be regarded as a further trial parameter of the variational method, to be determined by
its stationary value according to the method of minimal sensitivity\cite{minimal}.

The simple choice of an energy cutoff in the Euclidean space $p_E^2<\Lambda^2$ has the merit of
giving physical results that are directly comparable with lattice simulations where a finite lattice
acts just like an energy cutoff. Moreover, lattice simulations are the most natural benchmark for any
variational calculation in the low energy limit.
Thus we borrow from lattice simulation the regulating scheme and
its physical interpretation in terms of a bare interaction parameter $g=g(\Lambda)$ which is supposed to be
dependent on the energy scale $\Lambda$. Renormalization Group (RG) invariance requires that the physical
observables are left invariant by a change of scale $\Lambda\to\Lambda^\prime$ that is accompanied by
the corresponding change of the bare interaction $g(\Lambda)\to g(\Lambda^\prime)$.
Then, renormalized physical quantities can be defined that do not depend on the cutoff.
The theory has only one free parameter, namely the interaction strength $g$ at a given scale $\Lambda$,
or the scale $\Lambda$ at a given interaction strength. Once that is fixed, the function $g(\Lambda)$
can be determined by RG invariance. In lattice simulations, the scale $\Lambda$ is determined by a comparison
of some physical observables with their actual experimental value. In the present calculation we will limit
ourselves to the calculation of the propagators and we will fix the scale by a direct
comparison with the available lattice data.
It is important to point out that the present regularization scheme does not need the inclusion of any
counterterm in the Lagrangian and  especially mass counterterms that are forbidden by the gauge invariance of
the Lagrangian, but are sometimes included as free parameters.

Since $\Lambda$ is the unique energy scale in the theory, we will basically set $\Lambda=1$ and work in
units of the cutoff, at a given bare interaction strength $g$. Thus the choice of $\Lambda$ will be
equivalent to fixing the natural energy units. Any numerical solution of the stationary equations 
would take the form of a generic bare dressing function $f_B(x,g)$ where $x=p/\Lambda$ is the Euclidean
momentum in units of $\Lambda$ and $g=g(\Lambda)$. Since we only use the Euclidean formalism in this
section, we drop the $E$ in the momentum $p_E$ and denote by $p$ the Euclidean momentum unless
otherwise specified.
RG scaling requires that a renormalized function $f_R$
can be defined at an arbitrary scale $\mu$ by  {\it multiplicative} renormalization 
\BE
f_R({p}/{\mu}, \>\mu)=\frac{f_B(p/\Lambda,g)}{Z(g,\mu)}.
\label{Z}
\EE
For instance, as normalization condition we can require that $f_R=1$ at $p=\mu$ so that
\BE
Z(g,\mu)=f_B(\mu/\Lambda, g(\Lambda))
\EE
which is a function of $g$ and $\mu$ only, since $\Lambda$ can be regarded as an implicit function of $g$.
This kind of renormalization obviously requires that the dressing function shows the scaling property
\BE
f_R(p/\mu,\mu)=\frac{f_B(p/\Lambda, g(\Lambda))}{f_B(\mu/\Lambda, g(\Lambda))}=
\frac{f_B(p/\Lambda^\prime, g(\Lambda^\prime))}{f_B(\mu/\Lambda^\prime, g(\Lambda^\prime))}
\EE
or in other words the renormalized function $f_R$ is independent of $\Lambda$ and $g$ and
the bare function must satisfy
\BE
{f_B(p/\Lambda, g(\Lambda))}= K(g,g^\prime) \> {f_B(p/\Lambda^\prime, g(\Lambda^\prime))}
\label{scaling}
\EE
where $K(g,g^\prime)$ is a scaling constant that can only depend on $g$ and $g^\prime$.
This scaling property is evident in a log-log plot of the bare dressing functions since
the curves can be put one on top of the other by a simple shift of the axes. Since the approximation
and the numerical integration could spoil the scaling properties of the dressing functions, we will
consider the scaling as a test for the accuracy of the whole procedure.

\begin{figure}[t] \label{fig:Dp}
\centering
\includegraphics[width=0.35\textwidth,angle=-90]{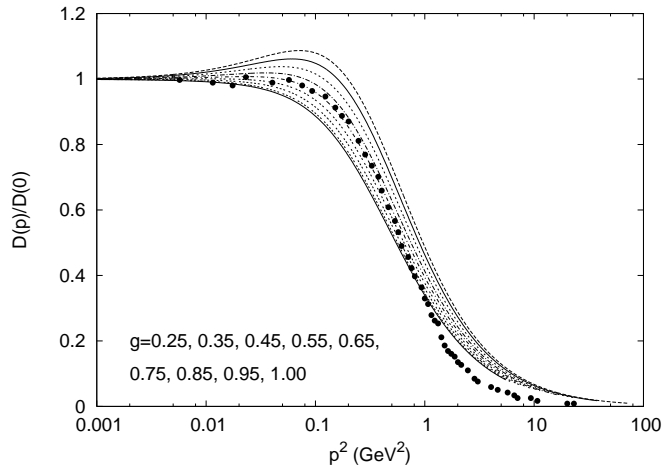}
\caption{The gluon propagator $D(p)/ D(0)$ as a function of the Euclidean momentum for several
values of the bare coupling $g=0.25, 0.35, 0.45, 0.55, 0.65, 0.75, 0.85, 0.95, 1$ (from the top
to the bottom). For each bare coupling the energy scale is fixed by taking $M=0.5$ GeV.
The Landau gauge lattice data of Ref.\cite{bogolubsky} ($g=1.02$, L=96) are reported as filled circles.}
\end{figure}

\begin{figure}[t] \label{fig:DRlog}
\centering
\includegraphics[width=0.35\textwidth,angle=-90]{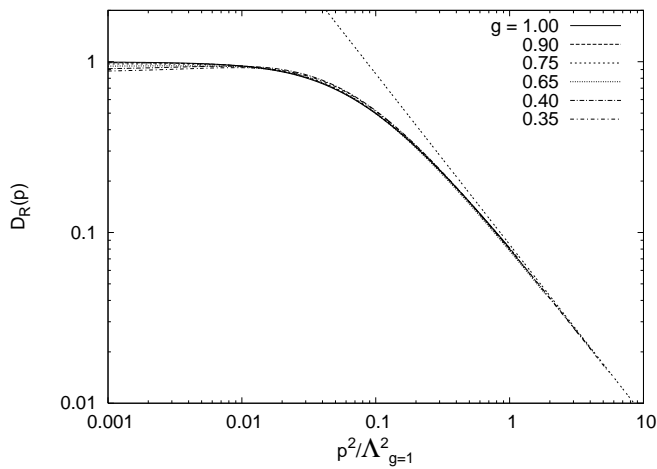}
\caption{Log-log plot of the renormalized propagator $D_R(p)$ as obtained by appropriate scaling of the bare propagator
for the bare coupling $g=0.35, 0.40, 0.65, 0.75, 0.90, 1$. The scale is arbitrary because of scaling:
all curves have been scaled in order to fall on top of the $g=1$ bare propagator of Fig.3.
Energy is in units of $\Lambda_{g=1}$ so that $g(1)=1$ (for $g=1$ the curve is not rescaled).
The dotted straight line is a fit of the asymptotic behavior by $D(p)\approx z/p^2$ and $z=0.085$.}
\end{figure}

As anticipated, in units of the cutoff, the proper polarization and self energy graphs in Fig.2 
are given by finite integrals and can be numerically evaluated as described in detail in the appendix,
making use of some initial choice for the trial dressing functions $f$ and $\chi$.
Inserting the actual value of the functions $\Pi_2^\star$ and $\Sigma_2^\star$ in Eqs.(\ref{dress}), 
a new pair of dressing
functions is obtained and the procedure can be iterated up to self consistency. Convergence is quite
fast and special normalization constraints can be imposed on the solutions by Eq.(\ref{Z}) or by
other boundary conditions.
The gluon propagator is reported in Fig.3  for several values of the 
bare coupling. For a rough comparison with lattice data, 
the energy scale is fixed by taking $M=0.5$ GeV in physical units. Actually, since the ratio $\tilde M=M/\Lambda$ 
is given by Eq.(\ref{gap}) at any coupling $g$, we are just taking $\Lambda=(0.5\>{\rm GeV})/(\tilde M)$.  
Lattice data from Ref.\cite{bogolubsky} are included
in the figure, but we must warn that the data of the simulation are obtained in the Landau gauge, while the
present calculation is in Feynman gauge. While the propagator is not expected to be gauge invariant, the
physical mass should not be too much sensitive to the gauge choice, and we may extract a rough estimate
of the energy scale by comparison of the data. Basically, in Fig.3
we are reporting $D(p)/ D(0)$ assuming that the mass parameter does not depend on the bare coupling $g$
and is kept fixed at the arbitrary value $M=0.5$ GeV that fits the data well enough. 
This is just a first estimate of the dynamical mass. A more
accurate estimate can be obtained by scaling, but depends on the actual definition of mass, which is not 
obvious as we will see later.

\begin{figure}[t] \label{fig:DR}
\centering
\includegraphics[width=0.35\textwidth,angle=-90]{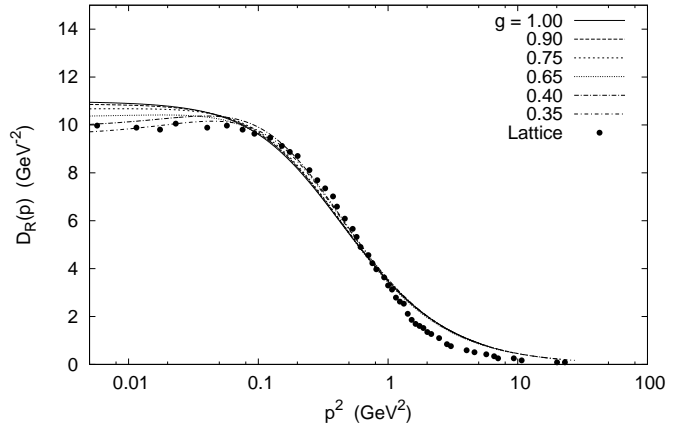}
\caption{The renormalized propagator $D_R(p)$ in physical units for the bare coupling 
$g=0.35, 0.40, 0.65, 0.75, 0.90, 1$. Scale factors are the same as in Fig.4 but the energy
scale has been fixed in order to fit the lattice data of Ref.\cite{bogolubsky} ($g=1.02$, L=96)
that are displayed as filled circles.}
\end{figure}

As shown in the log-log plot of Fig.4, by an appropriate change of scale the renormalized propagator $D_R(p)$ becomes
independent of $g$ and all the curve fall one on top of the other as expected from Eq.(\ref{scaling}). Here the
single curves are rescaled (just translated in the log-log plot) in order to fall on top of the $g=1$ bare propagator.
Scaling is rather good with the exception of the far infrared region.

The same curves have been reported in physical units and compared with the lattice data of Ref.\cite{bogolubsky}
in Fig.5. Scale ratios are the same as for Fig.4, but a physical energy scale is fixed in order to give a rough fit
of the lattice data. Despite the use of a different gauge, the main features of the lattice propagator seem to be 
reproduced by the trial function, with a pronounced flat behavior in the infrared. Here we are taking 
$\Lambda_{g=1}=2.24$ GeV and by using the scale factors that come from scaling, the mass parameter $M$ turns out
to be slightly dependent on $g$ with a value that goes from $M=0.47$ GeV at $g=1$ to $M=0.43$ GeV at $g=0.35$.
Thus using units of $M$, as we did in Fig.3, does not provides the best scaling. 

It is quite obvious that a slightly smaller energy scale in Fig.5
would give a better agreement in the UV region and a worsening in the IR, with a slight decrease of $M$ in physical
units. In other words the renormalized function $D_R(p)$ cannot be made to match the lattice data exactly 
over the whole range of $p$, yielding only approximate estimates of the mass parameter. 
That could be just a consequence of a different behavior of the propagator in different
gauges or it could be a shorthand of the variational approximation. Unfortunately we could not find any recent
lattice data in Feynman gauge to compare with.

\begin{figure}[b] \label{fig:mass}
\centering
\includegraphics[width=0.4\textwidth,angle=-90]{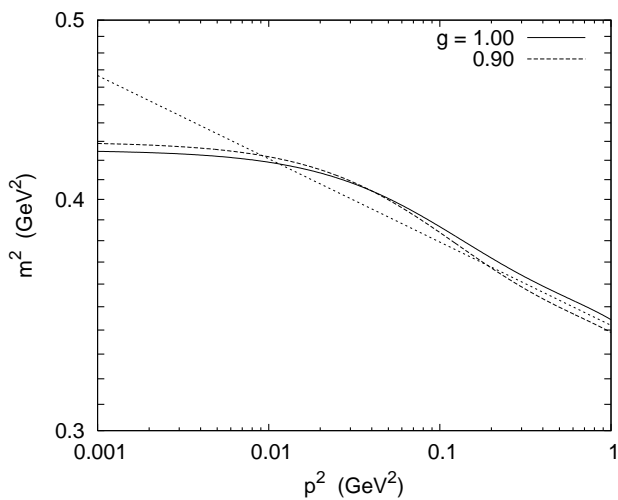}
\caption{The physical dynamical mass $m^2(p)$ is evaluated by Eq.(\ref{m2}) and displayed in a log-log plot for 
$g=0.9$ and  $1$. The leading behavior $m^2(p)\approx (p^2)^{-\eta}$ is shown by the dotted straight line for 
$\eta=0.045$.}
\end{figure}

Besides, it is not obvious that $M$ and $D$ should be the best estimates of the physical mass and propagator. 
The trial propagator $D$ is just the best zeroth order propagator that optimizes the convergence of the
perturbation expansion according to the method of stationary variance. That should give the best effective potential,
but the best two-point functions should be extracted as derivatives of the effective potential with respect to
the external field\cite{bubble}, and the result could be slightly different. 
That is probably the best way to approximate the true physical two-point functions in this framework, but
it requires the knowledge of the effective action as a functional of the external field, and we leave it as an
interesting further development. A more straightforward way to improve on the general behavior of the
propagator is provided by the second order approximation of Eq.(\ref{D2}), as we will discuss later 
in the next section.

We can define a dynamical physical mass $m(p)$ by requiring that
\BE
D(p)=\frac{z}{p^2+m^2(p)}
\label{m}
\EE
and that $m$ vanishes in the UV limit. The normalization constant $z$ can be extracted by the asymptotic
behavior $D(p)\approx z/p^2$ as shown in Fig.4. The mass then follows as
\BE
m^2(p)=\frac{z}{D(p)}-p^2
\label{m2}
\EE
provided that $p$ is not too large. In the far UV limit the two terms on the right hand side of Eq.(\ref{m2})
are very large and their difference gets very small, so that numerical errors make the result unreliable.
We show  a log-log plot of $m^2(p)$ in Fig.6 for two values of bare coupling $g=1, 0.9$. Deviations
from the exact scaling get amplified by this procedure, but nevertheless we may extract a power-law
behavior $m^2(p)\approx (p^2)^{-\eta}$ which is displayed as a straight dotted line in the log-log plot,
with $\eta=0.045$. This coefficient seems to be too small when compared to other predictions in Landau
gauge\cite{aguilar14}. The mass saturates in the IR limit at the value $m(0)\approx 0.65$ GeV which is slightly
larger than the mass parameter $M\approx 0.47$ GeV.

\section{Second order approximation}

The trial propagator $D(p)$, solution of the stationary conditions Eqs.(\ref{dress}), was
studied in the previous section as an optimized zeroth order approximation for the physical
gluon propagator. However, the dressing functions $f(p)$, $\chi(p)$ can be regarded as just
an optimal choice for the zeroth order starting point of the expansion, but they could have
no physical relevance. In other words the trial functions $D$, $G$ don't need to be the 
physical propagators but they could be regarded as just an infinite set of variational parameters. 
Of course, if the expansion is
optimized, the zeroth order functions must be close to the true physical propagators, and
that is the reason why we have traded them as good approximations for the physical propagators
until now.

A consistent way to evaluate the physical propagators would require the knowledge of the effective
action as a function of the external fields, in order to write the two-point functions as
functional derivatives of the effective action\cite{bubble}. We have left that task to future work.
However, it is quite reasonable to think that the actual approximation could also be  
improved by just adding higher-order Feynman graphs to the propagators $D$, $G$, 
using the same Feynman rules as we did 
before, with an optimized zeroth order action $S_0$ and an interaction $S_I$ that now are 
entirely specified by the knowledge of the dressing functions $f$, $\chi$. 

\begin{figure}[t] \label{fig:DB}
\centering
\includegraphics[width=0.35\textwidth,angle=-90]{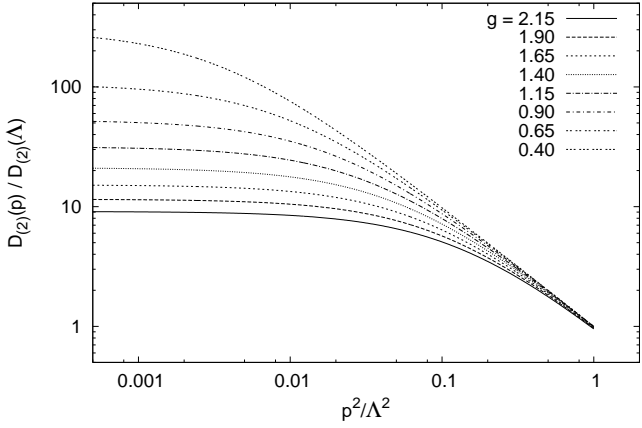}
\caption{The bare second order propagator $D_{(2)}(p)/D_{(2)}(\Lambda)$ is shown in a log-log plot for 
$g$ = 0.4, 0.65, 0.9, 1.15, 1.4, 1.65, 1.9, 2.15, in units of the cutoff $\Lambda$.}
\end{figure}

Thus taking $D_{(0)}(p)=D(p)$ and $G_{(0)}(p)=G(p)$ as the free propagators,
we can build the higher order functions $D_{(n)}$, $G_{(n)}$ by Dyson equations like Eq.(\ref{Dn}).
Dropping the color indices and the $E$ in the Euclidean momentum, 
the second order function in Eq.(\ref{D2}) can be written in the Euclidean formalism\footnote{In this Section,
as in the previous one, 
we drop the $E$ in the momentum $p_E$ and denote by $p$ the Euclidean momentum unless otherwise specified.}
\BE
{D_{(2)}^{-1}}_{\mu\nu}(p)=\eta_{\mu\nu}(p^2+M^2)-{\Pi_2^\star}_{\mu\nu}(p)
\label{D2M}
\EE
and by the same argument
\BE
{G_{(2)}^{-1}}(p)=G^{-1}(p)-\Sigma_1(p)-{\Sigma_2^\star}(p)
\EE
that by Eq.(\ref{S1}) becomes
\BE
{G_{(2)}^{-1}}(p)=\frac{-p^2}{\chi_{(2)}(p)} =- p^2-{\Sigma_2^\star}(p)
\label{G2}
\EE
where $\chi_{(2)}$ is a second order ghost dressing function that can be written as
\BE
{\chi_{(2)}(p)} =\left[{1+ \frac{{\Sigma_2^\star}(p)}{ p^2}}\right]^{-1}=
\frac{1}{ \chi(p)}.
\EE
It is remarkable that the second order ghost dressing function is just the reciprocal of the
zeroth order function $\chi$, thus we do not expect that both of them could give a reasonable
approximation for the true dressing function unless $\chi(p)$ is almost constant. That is indeed
the case, prompting to a decoupled scenario with ghosts that behave as free particles.

\begin{figure}[t] \label{fig:D2}
\centering
\includegraphics[width=0.35\textwidth,angle=-90]{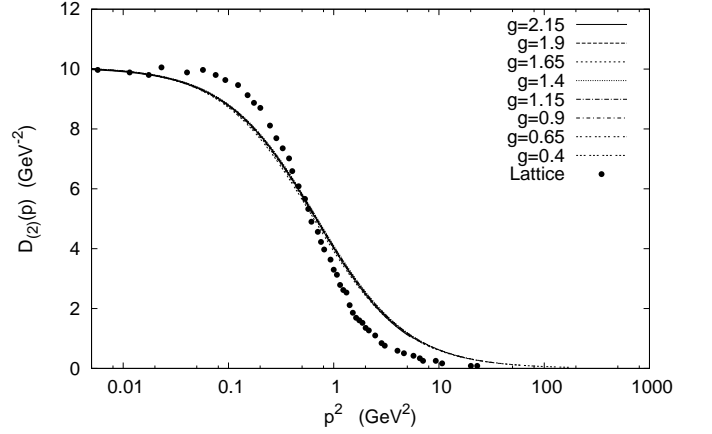}
\caption{The renormalized second order propagator is shown in physical units for the same 
couplings of Fig.7. By scaling, all the curves fall one on top of the other. The energy scale is
fixed by a rough fit of the Landau-gauge lattice data of Ref.\cite{bogolubsky} ($g=1.02$, L=96) 
that are dispayed as  filled circles.}
\end{figure}

By Lorentz invariance the gluon propagator can be written as
\BE
{D_{(2)}}_{\mu\nu} (p)=\eta_{\mu\nu} D_{(2)}(p) +p_\mu p_\nu D^{\prime\prime}_{(2)}(p)
\EE
where $D^{\prime\prime}$ is the longitudinal component, while $D_{(2)}$ is the physically relevant
part we are interested in.~\footnote{Actually, the function $D^{\prime\prime}$ is not relevant in the
calculation of any physical observable if the polarization is strictly transversal as required by gauge
invariance at any order of perturbation theory. In this calculation the polarization is only approximately
transversal.}
Since $\Pi_2^\star$ has the same Lorentz structure,
we only need the coefficient of $\eta_{\mu\nu}$, denoted by $\Pi_2^{\prime}$ in the appendix and
evaluated in detail by Eq.(\ref{Piprime}) for each of the 1PI graphs of Fig.2. 
The second order gluon propagator $D_{(2)}$ can then be written as 
\BE
\left[ D_{(2)}(p)\right]^{-1}=p^2+M^2-{\Pi_2^\star}^{\prime} (p).
\label{DB}
\EE

If we only retain the ghost-loop $\Pi^\prime_{2a}$ and neglect all other polarization graphs, the second-order
propagator becomes formally equivalent to that obtained by the curvature approximation of Ref.\cite{reinhardt14},
or by the improved GEP\cite{AF,HT} of Eqs.(\ref{curvature1}),(\ref{curvature2}). 
Actually, as shown by Eq.(\ref{curv}) in the appendix, the function $-\Pi_{2a}^\prime$ is
formally equivalent to the curvature of Ref.\cite{reinhardt14}, but the resulting propagator can be different
because of the dressing functions in the loop that have been calculated by different methods. For instance,
the ghost dressing function is set to its zeroth order in the GEP, it is given by the coupled stationary conditions 
of Eqs.(\ref{dress}) in the present method while is given by a closed set of self-consistent Dyson-Schwinger equations 
in Ref.\cite{reinhardt14}.

The second order gluon propagator that emerges from Eq.(\ref{DB}) seems to go a step forward by the
inclusion of all the 1PI graphs besides the ghost loop. For instance the gluon loop $\Pi^\prime_{2b}$ is
now summed together with the ghost loop $\Pi^\prime_{2a}$ as it should be for a correct cancellation
of the unphysical degrees of freedom.

The bare second-order propagator is shown in Fig.7 for several values of the bare coupling $g$. 
As for the trial function $D$, the log-log
plot prompts towards the existence of scaling properties . In fact, by a proper
scaling, all curves can be put one on top of the other, as displayed in Fig.8 where a physical energy scale has
been chosen in order to give a rough fit of the lattice data. 
While the scaling is now very good, it is painfully obvious that the agreement with the (Landau gauge) 
lattice data of Ref.\cite{bogolubsky} is very poor
and only a very loose energy scale can be fixed by this method. Once more, we expect that relevant differences
may exist between propagators in different gauges and these differences may also depend on the formal
definition of the propagator that is not an observable quantity but just an intermediate scheme-dependent 
step of the full calculation. Our choice for the scale $\Lambda^2$ could become $3\Lambda^2$ or $0.3\Lambda^2$ if
we would like to improve the agreement in the IR or in the UV respectively.
Any prediction for the dynamical mass also depends on the way the mass is defined, and we have already met different
estimates in the study of the trial propagator $D$ in the previous section.

\begin{figure}[t] \label{fig:m2}
\centering
\includegraphics[width=0.35\textwidth,angle=-90]{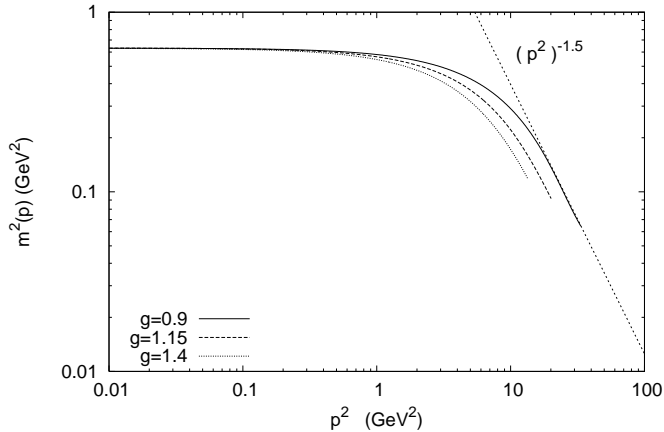}
\caption{The second-order dynamical mass of Eq.(\ref{mB}) is reported in a log-log plot with the
same scaling factors and energy scale of Fig.8, for $g$ = 0.9, 1.15 and 1.4. The straight dotted line
shows the power behavior $\sim (p^2)^{-\eta}$ with the exponent $\eta=1.5$.}
\end{figure}

The renormalized second order propagator $D_{(2)} (p)$ in Fig.8 can be fitted quite well by the simple expression
\BE
D_{(2)}(p)\approx\frac{Z}{p^2+m^2}
\label{fit1}
\EE
yielding a physical mass parameter $m\approx 0.8$ GeV that is basically independent of $g$.
Of course our uncertainty on the energy scale would give something like $0.4{\>\rm GeV}< m < 1.4 {\>\rm GeV}$.

We can introduce a better definition for the dynamical mass if we take
\BE
D_{(2)}(p)=\frac{Z}{p^2+ m^2(p)}
\label{fit2}
\EE
where now $m(p)$ is a function which is supposed to decrease as a power, 
$m^2\sim (p^2)^{-\eta}$ for large energies.
While Eq.(\ref{fit2}) is just a definition for $m$, by Eq.(\ref{DB}) it can be written as
\BE
m^2(p)= p^2(Z-1)+Z\left(M^2-{\Pi_2^\star}^\prime (p)\right)\sim 
\left(\frac{p^2}{p_0^2}\right)^{-\eta}.
\label{mB}
\EE
The parameter $Z$ can be tuned in order to get a power-law behavior  that would
appear as a linear curve in a log-log plot. 
For any bare coupling we find $Z\approx 1$, as we expected by the knowledge of the exact asymptotic limit.
For instance, in the case of $g=0.9$ the asymptotic behavior of $m(p)$ is fitted by $Z=0.9978$.
The exponent turns out to be $\eta=1.5$ for any coupling, as shown in Fig.9 where the function $m^2(p)$ is
reported for some different values of the bare coupling, with the same scaling factors and energy scale
of Fig.8. Scale dependent values of $\eta$, oscillating in the range $1.08<\eta<1.26$, 
have been reported in Landau gauge by Ref.\cite{aguilar14}.
While this kind of plot enhances minor deviations from the exact scaling, we find the same high-energy
power-law behavior for different couplings, with a dynamical mass $m(p)$ that saturates 
at $m(0)\approx 0.8$ GeV.

Since we have seen that $Z\approx 1$, the second-order propagator can be written as $D_{(2)}^{-1}=p^2+m^2(p)$
with a dynamical mass that takes the simple form
\BE
m^2(p)=M^2-{\Pi_2^\star}^\prime (p),
\label{mB2}
\EE
suggesting that the full polarization matrix could be written as the
sum of a constant shift plus a transverse polarization term
\BE
{\Pi_2^\star}_{\mu\nu} (p)=-(\delta m^2)\eta_{\mu\nu}-
\pi(p)\left(\eta_{\mu\nu}-\frac{p_\mu p_\nu}{p^2}\right)
\label{trans}
\EE
where $\delta m^2=m^2(0)-M^2$ is a constant second-order mass shift and $\pi (p)$, 
the coefficient of the transverse part, must vanish in the low energy
limit, $\pi(0)=0$, yielding
\BE
m^2(p)=m^2(0)+\pi(p).
\label{mB3}
\EE

While Eq.(\ref{trans}) is suggestive, it is not obvious in any way that it should
hold since it requires that
\BE
-\pi(p)=\left[{\Pi_2^\star}^\prime (p)-{\Pi_2^\star}^\prime (0)\right]
=-{\Pi_2^\star}^{\prime\prime} (p)
\label{trans2}
\EE
having denoted by ${\Pi_2^\star}^\prime$ and ${\Pi_2^\star}^{\prime\prime}$ the coefficents
of $\eta_{\mu\nu}$ and $p_\mu p_\nu/{p^2}$ respectively, in the proper second-order
polarization function, with the notation of Eq.(\ref{pol}). 
These functions are defined in detail in the appendix in Eqs.(\ref{Pi}), (\ref{Piprime}).
The transversality (up to a constant) of the polarization function is what we would expect by gauge invariance
in presence of a dynamical mass. It is not obvious that it should hold in the present approximate scheme,
and it is generally achieved by a correct cancellation of the unphysical degrees of freedom by the ghost loops.
Thus, it is remarkable that Eqs.(\ref{trans}),(\ref{trans2}) hold, albeit approximately, in the present variational
calculation.

\begin{figure}[b] \label{fig:2a2b}
\centering
\includegraphics[width=0.35\textwidth,angle=-90]{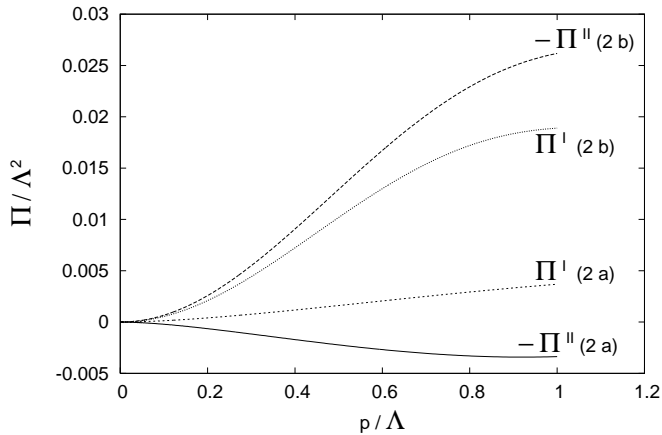}
\caption{The polarization functions $\Pi^\prime$, $-\Pi^{\prime\prime}$ for the ghost-loop graph $(2a)$ and 
the gluon-loop graph (2b) are displayed for a bare coupling $g=1.2$, in units of the cutoff. The functions $\Pi^\prime$
have been shifted by a constant in order to have $\Pi^\prime (0)=0$. The transversality condition
of Eq.(\ref{trans2}), requiring that $\Pi^\prime\approx -\Pi^{\prime\prime}$, is not satisfied by the single terms.}
\end{figure}

\begin{figure}[t] \label{fig:Ptot}
\centering
\includegraphics[width=0.35\textwidth,angle=-90]{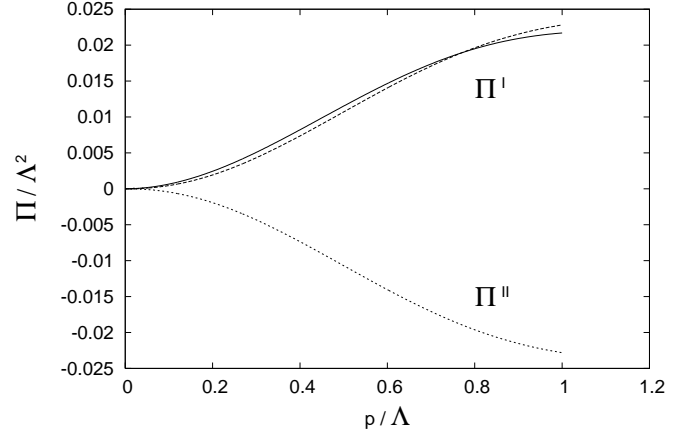}
\caption{Functions $\Pi^\prime$, $\Pi^{\prime\prime}$ for the total polarization function, including
all the 1PI second-order graphs of Fig.2. The function $\Pi^\prime$ is shifted by the constant term $\delta m^2$
in order to have $\Pi^\prime (0)=0$. 
The opposite of $\Pi^{\prime\prime}$, namely $-\Pi^{\prime\prime}$, is also shown as a dotted line for
a direct comparison. We observe that $\Pi^\prime\approx -\Pi^{\prime\prime}$, and the transversality
condition Eq.(\ref{trans2}) is approximately satisfied when all graphs are added together.}
\end{figure}

\begin{figure}[b] \label{fig:chi}
\centering
\includegraphics[width=0.35\textwidth,angle=-90]{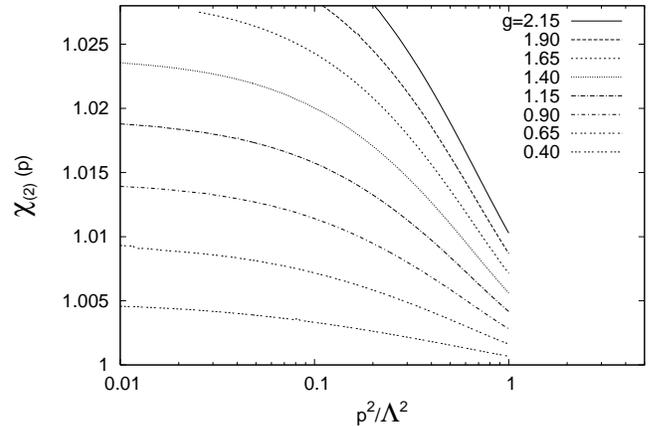}
\caption{The bare second-order ghost dressing function
$ \chi_{(2)}$ for several values of the bare coupling $g$, in units of the cutoff.} 
\end{figure}

It is instructive to go in detail and look at the behavior of the single terms of the polarization. As expected,
Eq.(\ref{trans2}) is not satisfied by the single graphs of Fig.2. For instance, in Fig.10,
for the ghost loop ${\Pi_{2a}}^{\mu\nu}$ and the gluon loop ${\Pi_{2b}}^{\mu\nu}$, the
functions $\Pi^\prime$ and $-\Pi^{\prime\prime}$  are displayed at a bare coupling $g=1.2$, in units of the cutoff. 
The functions $\Pi^\prime$ have been shifted by a constant in order to have $\Pi^\prime (0)=0$ for
all the single terms. We observe that the transversality condition of Eq.(\ref{trans2}) would
require that the shifted functions should satisfy $\Pi^\prime\approx -\Pi^{\prime\prime}$ at least. 
That is not the case in Fig.10.

Now let us look at the functions $\Pi^\prime$, $\Pi^{\prime\prime}$ for the total polarization function, including
all the 1PI second-order graphs of Fig.2. The function $\Pi^\prime$, now shifted by the constant term $\delta m^2$, 
is shown in Fig.11 and compared to  $\Pi^{\prime\prime}$. 
Albeit approximately, $\Pi^\prime\approx -\Pi^{\prime\prime}$ when all graphs are added together.
Actually, the ghost and gluon loops almost satisfy the transversality condition when added together, but the
accuracy improves when all the graphs are summed. Since the loops are evaluated in terms of the self-consistent
dressing functions, the transversality condition Eq.(\ref{trans2}) seems to be an important test for the
overall reliability of the calculation.

Finally, a note on the ghost dressing functions is in order. Since ghosts are not physical, their properties
are expected to be even more sensitive to a change of gauge. While the gluon mass can have a physical meaning,
no physical observable can be clearly related to ghosts. In fact ghosts can also disappear in a convenient gauge.
In Feynman gauge, even a finite ghost propagator, with a vanishing dressing function in the IR limit, 
would be still consistent with a finite gluon propagator as shown by a very recent study\cite{aguilarFeynman}.
Thus we have no clear way to say how accurate our ghost dressing function is and any comparison with
Landau gauge lattice data seems to be even more questionable than it already was for gluons.
We can just say that the ghost dressing function seems to play well its role of canceling the unphysical 
degrees of freedom, as discussed before, and that should suffice.
Both $\chi$ and $\chi_{(2)}$ are rather flat functions, with a decoupling scenario of almost free
ghosts and an IR finite gluon propagator. We do not see any evidence of a vanishing of the dressing function
in the IR limit.
In order to give a full picture, the bare second-order ghost dressing function
$ \chi_{(2)}$ is displayed in Fig.12 for several values of the bare coupling $g$. 
Despite the scale, the curves are very flat as compared to Landau gauge lattice data.

\section{Discussion}

One of the major achievements of the present paper is the proof that a physically consistent
solution does exist for the coupled set of non-linear integral equations that arise from the
condition of stationary variance. Since  pure Yang-Mills theory already contains fermions 
(the ghosts),  inclusion of quarks in the formalism is straightforward, and would open the way to
a broader study of QCD by the same method. 

Feynman gauge is also interesting by itself, because the IR behavior of the theory
is basically unexplored yet in that gauge.
The general picture that emerges from the calculation confirms 
the decoupling scenario, with a finite ghost dressing function, a
finite gluon propagator in the IR limit, and a dynamical mass
that decreases as a power in the UV limit. Any quantitative estimate
of the gluon mass requires that an accurate energy scale should be
fixed first: without any lattice data available in Feynman gauge,
we can only give a roughly approximate guess of the scale.
By comparison with Landau-gauge lattice data, a saturating value
$m(0)\approx 0.5 - 0.8$ GeV is found, depending on the precise
definition of mass and energy scale. That estimate is in agreement with
other predictions in Feynman gauge\cite{aguilarmass}.
From a qualitative point of view, we cannot confirm the prediction
of a finite ghost propagator that has been recently argued\cite{aguilarFeynman}.
The ghost propagator diverges in the IR limit, and the ghost behaves like 
a free zero-mass particle.

It is an open question if the second order approximation of Section VI really improves 
the gluon propagator. By a comparison with Landau-gauge lattice data the optimal trial function
$D(p)$ reproduces the flat IR behavior quite well in Fig.5, while the agreement worsen for the
second order function $D_{(2)}$ in Fig.8. Of course, we cannot take too seriously a
comparison between different gauges, 
but in principle the second order approximation could even spoil the variational result.
If we trust the comparison of the optimal trial function $D(p)$ with the Landau gauge data, 
than a bit more accurate estimate of the scale can be done, and the dynamical mass
saturates at a smaller value $m(0)\approx 0.65$ GeV as shown in Fig.6.
The functions $D(p)$ and $D_{(2)}(p)$ can be regarded as different approximations, and
for that reason we studied in detail the features of both of them in Section V and VI respectively.
Needless to say, we need some lattice data in Feynman gauge for answering the open question.

The method can be improved in many way. We did not bother about gauge invariance in this first
approach, but the properties of the polarization function, namely the correct cancellations of
the unphysical degrees of freedom by the ghosts, show that the constraints of gauge invariance
can be satisfied, at least approximately, by the variational solution.
While some attempts could be made for enforcing gauge invariance\cite{kogan,aguilar10}, a physically
motivated choice for the gauge would probably improve the approximation. Landau gauge would be a
good candidate, as it would enforce the transversality in the polarization function from the beginning.
An other interesting further development would come from the extension of the formalism to the general
case of a finite external background field. For a scalar theory that kind of approach allows a consistent
definition of approximate vertex functions by the functional derivative of the
effective action. For the GEP these functions can be shown to be the sum of an infinite set of 
bubble graphs\cite{bubble}. A similar approach would give a more consistent approximation for the gluon propagator
in the present variational framework.
Eventually, the inclusion of quarks would lead to a direct comparison with the low energy phenomenology
of QCD.

\appendix

\section{Explicit evaluation of the graphs and numerical details}

We give explicit integral representations of the 1PI graphs that are displayed in Fig.2. 
The graphs are evaluated by standard Feynman rules, with the trial functions $iG$ and $iD$ 
associated to any internal ghost and gluon line respectively, and the standard QCD vertices
that can be read from the Lagrangian terms  in Eqs.(\ref{Lint}). 

On general grounds, by Lorentz invariance and color symmetry, any generic term contributing to
the gluon polarization function can be written as 
\BE
\Pi_{\mu\nu}^{ab} (p)=\delta_{ab}\left[\eta_{\mu\nu} \Pi^\prime(p)
+\frac{p_\mu p_\nu}{p^2}\Pi^{\prime\prime} (p)\right].
\label{pol}
\EE
The functions $\Pi^\prime$ and $\Pi^{\prime\prime}$ can be extracted by saturating the indices
with different choices for the tensor $t_{\mu\nu}$ in Eq.(\ref{Psum}).
Taking $t_{\mu\nu}=\eta_{\mu\nu}$ Eq.(\ref{Psum}) yields
\BE
\Pi (p)= \frac{1}{32} \sum_{ab, \mu\nu} \delta_{ab} \eta^{\mu\nu} \Pi_{\mu\nu}^{ab}(p)=
\Pi^\prime (p) + \frac{1}{4}\Pi^{\prime\prime} (p)
\label{Pi}
\EE
while taking $t_{\mu\nu}(p)=\eta_{\mu\nu}-p_\mu p_\nu/p^2$ 
\BE
\Pi^\prime (p)= \frac{1}{24} \sum_{ab, \mu\nu} \delta_{ab} \left(
\eta_{\mu\nu}-\frac{p_\mu p_\nu}{p^2} \right) \Pi_{\mu\nu}^{ab}(p).
\label{Piprime}
\EE
The function $\Pi^{\prime\prime}$ follows as $\Pi^{\prime\prime}=4(\Pi-\Pi^\prime)$.

The function $\Pi$ is the one required for the stationary equations and must be inserted
in Eqs.(\ref{dress}) for the evaluation of the self consistent solution.
The function $\Pi^\prime$ has been used for the evaluation of the physically relevant part
of the propagator at higher orders in Eq.(\ref{DB}). 

The numerical integration has been performed by successive one-dimensional integrations by
the standard Simpson method in the Euclidean space and with an energy cutoff $p_E^2<\Lambda^2$.  
Four-dimensional integrals of simple functions of $k^2_E$
are reduced to simple one-dimensional integrals before numerical integration, according to
\BE
\int_\Lambda \kkkE A(k^2_E)=\frac{1}{8\pi^2}\int_0^\Lambda A(k^2) k^3 {\rm d}k.
\EE
Four-dimensional integrals of functions of the two variables $(k_E\cdot p_E)$ and $k^2_E$ are reduced 
to two-dimensional integrals according to

\begin{align}
\int_\Lambda \kkkE &A[(k_E \cdot p_E), k^2_E]=\nn\\
&=\int_0^\Lambda\frac{y^2 {\rm d}y}{4\pi^3}
\int_{-\sqrt{\Lambda^2-y^2}} ^{\sqrt{\Lambda^2-y^2}} A[(xp_E), (x^2+y^2) ] {\rm d}x.
\label{2dim}
\end{align}

\begin{widetext}
\subsection{Graph (2a)}
The ghost loop $\Pi_{2a}$ can be written as 
\BE
{\Pi_{(2a)}}^{cd}_{\mu\nu}(p)=-ig^2
f_{abc}f_{bad}\int\kkk iG(p+k)iG(k) (p_\mu+k_\mu)k_\nu=-\delta_{cd} Ng^2
\int\kkki\frac{(p_\mu+k_\mu)k_\nu}{(p+k)^2k^2} \chi(p+k)\chi(k)
\EE
where a minus sign has been inserted because of the fermion loop. We can saturate the indices
as shown in Eqs.(\ref{Pi}),(\ref{Piprime}) and write in the Euclidean space
\BE
\Pi_{2a} (p_E)= -\frac{Ng^2}{4} \int\kkkE \frac{\chi(p_E+k_E)\chi(k_E)}{(p_E+k_E)^2k_E^2}
\left( p_E\cdot k_E +k_E^2\right)
\EE
\BE
\Pi^{\prime}_{2a} (p_E)= -\frac{Ng^2}{3} \int\kkkE \frac{\chi(p_E+k_E)\chi(k_E)}{(p_E+k_E)^2}
\left(1-\frac{(p_E\cdot k_E)^2}{p_E^2 k_E^2}\right). 
\label{curv}
\EE
We observe that  the function $-\Pi^{\prime}_{2a}$ is the curvature 
function of Ref.\cite{reinhardt14}  as expected by Eq.(\ref{curvature2}) or Eq.(\ref{D2}).
Actually $\Pi^\prime_{2a}$ is the correct function that must be inserted in Eqs.(\ref{curvature2}), 
(\ref{D2}) in order to extract the physically relevant part of the gluon propagator.

\subsection{Graph (2b)}
The gluon loop $\Pi_{2b}$ can be written as 
\begin{align}
-i{\Pi_{(2b)}}^{ad}_{\mu\nu}(p)=\frac{g^2}{2}
f_{abc}f_{dbc}\int\kkk iD(p+k)iD(k)\>&
\left\{(2p_\tau+k_\tau)\eta_{\mu\rho}-(2k_\mu+p_\mu)\eta_{\rho\tau}+(k_\rho-p_\rho)\eta_{\mu\tau}\right\}\times\nn\\
&\>\>\times\left\{(p_\nu+2k_\nu)\eta^{\rho\tau}-(k^\tau+2p^\tau)\eta_\nu^\rho+(p^\rho-k^\rho)\eta_\nu^\tau\right\}
\end{align}
where a symmetry factor $1/2$ has been inserted. By trivial algebra
\BE
{\Pi_{(2b)}}^{ad}_{\mu\nu}(p)=\delta_{ad}\frac{Ng^2}{2}
\int\kkki D(p+k)D(k)
\left\{\eta_{\mu\nu}(5p^2+2k^2+2pk)+(10k_\mu k_\nu-2p_\mu p_\nu+5k_\mu p_\nu+5p_\mu k_\nu)\right\}
\EE
and then taking $t_{\mu\nu}=\eta_{\mu\nu}$, Eq.(\ref{Psum}) yields
\BE
\Pi_{2b} (p_E)= \frac{9Ng^2}{4} \int\kkkE \frac{f(p_E+k_E) f(k_E)}{(p_E+k_E)^2k_E^2}
\left(p_E^2+p_E\cdot k_E +k_E^2\right),
\EE
while Eq.(\ref{Piprime}) reads
\BE
\Pi^\prime_{2b} (p_E)= \frac{Ng^2}{2} \int\kkkE \frac{f(p_E+k_E) f(k_E)}{(p_E+k_E)^2k_E^2}
\left[5p_E^2+2p_E\cdot k_E +2k_E^2
+\frac{10}{3}\left(k_E^2-\frac{(p_E\cdot k_E)^2}{p_E^2}\right)
\right].
\EE

\subsection{Graph (2c)}
Let us denote by $\Gamma_{abcd}^{\mu\nu\rho\sigma}$ the four-gluon vertex ${\cal L}_4$
in Eq.(\ref{Lint}), that can be written as
\BE
\Gamma^{abcd}_{\mu\nu\rho\sigma}=-i\frac{g^2}{4!}\left[T^{abcd}_{\mu\nu\rho\sigma}+
T^{acdb}_{\mu\rho\sigma\nu}+T^{adbc}_{\mu\sigma\nu\rho}\right]
\label{Gamma4}
\EE
where the matrix structure $T$ is 
\BE
T^{abcd}_{\mu\nu\rho\sigma}=f_{eab}f_{ecd}(\eta_{\mu\rho}\eta_{\nu\sigma}-\eta_{\mu\sigma}\eta_{\nu\rho}).
\EE
With a symmetry factor $(4!4!/3!)$, the two-loop term $\Pi_{2c}$ can be written as
\BE
-i{{\Pi_{(2c)}}^{af}}_{\mu\tau}(p)= \frac{4!4!}{3!} 
{\Gamma^{abcd}}_{\mu\nu\rho\sigma}{ {\Gamma^{fbcd}}_\tau}^{\nu\rho\sigma}
\int\kkk\int\qqq iD(k)iD(q)iD(k+q+p)
\EE
and in terms of the matrix structure $T$
\BE
{{\Pi_{(2c)}}^{af}}_{\mu\tau}(p)= 3\frac{g^4}{3!} 
T^{abcd}_{\mu\nu\rho\sigma}\left[{{T^{fbcd}}_\tau}^{\nu\rho\sigma}+
{{T^{fcdb}}_\tau}^{\rho\sigma\nu}+{{T^{fdbc}}_\tau}^{\sigma\nu\rho}\right]
\int\kkki\int\qqqi D(k)D(q)D(k+q+p)
\EE
where the  $3$ factor in front arises because of the three identical terms in the product $\Gamma\cdot\Gamma$
that only differ for a permutation of dummy indices.
The first product is
\BE
T^{abcd}_{\mu\nu\rho\sigma}{{T^{fbcd}}_\tau}^{\nu\rho\sigma}=2(\eta_{\mu\tau}\eta^{\nu}_{\nu}-\eta_{\mu\tau})
f_{eab}f_{ecd}f_{gfb}f_{gcd}=6\eta_{\mu\tau}
N\delta_{eg} f_{eab}f_{gfb}=6N^2\delta_{af}\eta_{\mu\tau}
\EE
while the other two products are
\BE
T^{abcd}_{\mu\nu\rho\sigma}\left[{{T^{fcdb}}_\tau}^{\rho\sigma\nu}+{{T^{fdbc}}_\tau}^{\sigma\nu\rho}\right]=
-(\eta_{\mu\tau}\eta^{\nu}_{\nu}-\eta_{\mu\tau})\left[f_{eab}f_{ecd}(f_{gfc}f_{gdb}+f_{gfd}f_{gbc})\right]=
3\eta_{\mu\tau}f_{eab}f_{ecd}(f_{gcd}f_{gfb})
\EE
having used Jacobi identity in the last equality. The last two lines can be summed together yielding
$9N^2\delta_{af}\eta_{\mu\tau}$ and by Eqs.(\ref{Pi}),(\ref{Piprime}) we obtain
\BE
\Pi_{2c}(p_E)=\Pi^\prime_{2c}(p_E)= \frac{9g^4N^2}{2} 
\int\kkkE\int\qqqE \frac{f(k_E)f(q_E)f(k_E+q_E+p_E)}{k_E^2q_E^2(k_E+q_E+p_E)^2}.
\EE
Before numerical integration, this  eight-dimensional integral is reduced to a four-dimensional one by obvious
generalization of Eq.(\ref{2dim}): the internal integration is performed on the two variables
$q_E^2$, $[q_E\cdot(k_E+p_E)]$ by Eq.(\ref{2dim}) and the resulting function of $k_E$ and $(k_E\cdot p_E)$ is
integrated again by Eq.(\ref{2dim}).

\subsection{Constant graphs (1b), (2d) and (2e)}
With a symmetry factor $4!/2$, the one-loop first order polarization $\Pi_{1b}$ can be written in terms of the
four-gluon vertex of Eq.(\ref{Gamma4})
\BE
-i{{\Pi_{(1b)}}^{cd}}_{\rho\sigma}= \frac{4!}{2} 
{\Gamma^{abcd}}_{\mu\nu\rho\sigma}
\int\kkk \left[i\eta_{\mu\nu}\delta_{ab}\right] D(k)=-i\frac{g^2}{2} 
\>(6N\delta_{cd}\>\eta_{\rho\sigma})\int\kkki D(k)
\label{1b}
\EE
yielding the result of Eq.(\ref{P1}) and $\Pi_{1b}=- M^2$ as defined in Eq.(\ref{gap}).

Since the total first order polarization $\Pi_1$ is diagonal, we can evaluate the second order terms
$\Pi_{2d}$ and $\Pi_{2e}$ directly from the first-order one-loop term $\Pi_{1b}$ by replacing the
internal gluon propagator $D$ with the product $D\Pi_1 D$ in Eq.(\ref{1b}), yielding
\BE
\Pi_{2d}= 3Ng^2 \int\kkki D(k)\left[ D^{-1}(k)- D_0^{-1} (k)\right]  D(k)=
3Ng^2\left[ I_0^{(2)}-I_0^{(1)}\right]
\label{2d}
\EE
\BE
\Pi_{2e}= 3Ng^2 \int\kkki D(k)\left[ - M^2 \right] D(k)=3Ng^2 M^2 I_1^{(2)}= M^4
\frac{I_1^{(2)}}{I_0^{(1)}}.
\label{2e}
\EE

\subsection{One-loop ghost self-energy}
The second-order proper self energy graph (third graph in Fig.2) can be written as
\BE
{\Sigma^\star_2}^{ad} (p)=g^2
f_{cba}f_{cdb}\int\kkki (-k\cdot p) iD(p+k)iG(k)
\EE
and switching to the Euclidean space
\BE
{\Sigma^\star_2}^{ad} (p_E)=\delta_{ad} \left[ 
Ng^2 \int\kkkE\frac{f(k_E+p_E)\chi(k_E)}{(k_E+p_E)^2 k_E^2}(k_E\cdot p_E)
\right].
\EE

\end{widetext}


\begin{thebibliography} {99}
\bibitem{aguilarmass} A collection of gluon mass estimates can be found in Table I of Ref.\cite{aguilar14}.
\end{thebibliography}
\end{document}